# Foundations of Information Theory

**M. Burgin**

Department of Mathematics
University of California, Los Angeles
405 Hilgard Ave.
Los Angeles, CA 90095

**Abstract.**

Information has become the most precious resource of society. At the same time, there is no consensus on the meaning of the term "information," and many researchers have considered problems of information definition. This results in a quantity of contradictions, misconceptions, and paradoxes related to the world of information. To remedy the situation, a new approach in information theory, which is called the general theory of information, is developed. The main achievement of the general theory of information is explication of a relevant and adequate definition of information. This theory is built on an axiomatic base as a system of two classes of principles and their consequences. The first class consists of the ontological principles, which are revealing general properties and regularities of information and its functioning. Principles from the second class explain how to measure information.

**Key words:** information, ontology, principles, infological system, structure, knowledge

## 1. Introduction.

Foundations of a scientific discipline is a systematic analysis of the most basic or fundamental concepts of this scientific discipline, that results in demonstration of fundamentality of the considered concepts, building theoretical (often mathematical) models for the concepts, finding properties of these concepts, and establishing the rules of operation with these concepts. For instance, foundations of mathematics studies structures that are used to build the whole mathematics, their properties and operations with them. There are set-theoretical, named-set-theoretical, categorical, algorithmic, and logical foundations of mathematics.

For information theory, the most basic is the concept of information. However, kinds and types of information and its theoretical representations form an extensive diversity of phenomena, concepts, formulas, and ideas. This inspired many researchers to argue that it is impossible to develop a unified definition of information (Belkin, 1978; Belkin and Robertson, 1976). In particular, the most prominent researcher in the field of information science, Claude Shannon, wrote (cf., (Shannon, 1993)) that it was hardly to be expected that a single concept of information would satisfactorily account for the numerous possible applications of the general field of information theory.

A persuasive argument for impossibility to give such a unified definition is given in (Capurro, Fleissner and Hofkirchner, 1999). This result and many other arguments (cf., for example, (Melik-Gaikazyan, 1997)) undermine generality of conventional definitions of information and imply impossibility of a universal definition of information.

At the same time, information has become the leading force in contemporary society (Bell, 1980) and rational development of information economy needs sound foundations, including a relevant comprising definition of information (Arrow, 1979; 1984; Godin, 2008). One more problem is to find adequate relations between data, knowledge, and information (Ackoff, 1989).

Nevertheless, it has become possible to synthesize all directions and approaches in information studies and to find a solution to the important problem of understanding what information is in the general theory of information. This was achieved in the general theory of information (Burgin, 1994; 1995; 2001; 2002; 2003; 2004) through utilization of a new definition type. Namely, to overcome limitations of the conventional approaches and to solve the problem of information definition a parametric definition is used in the general theory of information. Parametric systems (parametric curves, equations, functions, etc.) are frequently used in mathematics and its applications. For instance, a parametric curve in a plane is defined by two functions *f*(*t*) and *g*(*t*), while a parametric curve in space has the following form: (*f*(*t*), *g*(*t*), *h*(*t*) ) where parameter *t* takes values in some interval of real numbers.

Parameters used in mathematics and science are, as a rule, only numerical and are considered as quantities that define certain characteristics of systems. For instance, in probability theory, the normal distribution has as parameters the mean $\mu$ and the standard deviation $\sigma$. A more general parameter, functional, is utilized for constructing families of non-Diophantine arithmetics (Burgin, 1997a; 2001a).

In the case of the general theory of information, the parameter is even more general. The parametric definition of information utilizes a system parameter. Namely, an infological system plays role of a parameter that discerns different kinds of information, e.g., social, personal, chemical, biological, genetic, or cognitive, and combines all of the in one general concept "information".

The general theory of information is based on a system of principles. There are two groups of such principles: ontological and axiological. These principles single out what is information describing its properties, and thus, form foundations for information theory.

## 2. Ontological Principles of Information Theory

The main question here is "What is information?" To answer this question, we start with describing the basic properties of information in the form of ontological principles.

**Ontological Principle O1 (the Locality Principle).** *It is necessary to separate information in general from an information (or a portion of information) for a system **R**. In other words, empirically, it is possible to speak only about information (or a portion of information) for a system.*

Why is this principle so important? The reason is that all conventional theories of information assume that information exists as something absolute like time in the Newtonian dynamics. Consequently, it is assumed that this absolute information may be measured, used, and transmitted. On the abstract level, it is possible to build such a mathematical model that makes sense of absolute information, but in practical environment, or as scientists say, empirically, this is not so.

To demonstrate this, let us consider the following situation. We have a book in Japanese and want to ask what information it contains. For a person who does not know Japanese, it contains no information. At the same time, its information for those who know Japanese may be immense.

Another situation: let us consider a textbook, for example, in mathematics. If it is a good textbook, then it contains a lot of information for a mathematics student. However, if we show this book to a professional mathematician, she or he might say, "Oh, I know everything in this book, so it contains no information for me."

We will have the same result but for a different reason if we give this book to an art student who is bored with mathematics.

To make situation more evident, imagine a completely deaf and blind person who comes to a movie theater without any devices to compensate his deficiencies. How much information this person will get there?

It is interesting that the Ontological Principle O1 demonstrates tendencies and changes similar to those that were prevalent in theoretical physics in the $20^{th}$ century. Classical Newtonian-Laplacian physics is global, that is, all is the same whatever place in the universe we take. New physics has developed more refined methods. Relativity theory states that inertial systems that move with different speeds have different time. Quantum electrodynamics models quantum phenomena by gauge fields, which are invariant with respect to local transformations.

**Definition 1.** The system *R* with respect to which some information is considered is called the *receiver*, *receptor* or *recipient* of this information.

Such a receiver/recipient can be a person, community, class of students, audience in a theater, animal, bird, fish, computer, network, database and so on. Necessity to have a receiver stated in the Ontological Principle O1 implies, as Buckland (1991) explains, "that the capability of being informative, the essential characteristic of information-as-thing, must also be *situational*". In this context, to be informative for people means to have an answer to somebody's question. The *informativeness*, in the sense of Buckland (1991), is a relation between the question and the thing. Thus, there is no such a thing that is inherently informative. To consider something as information for an individual or group of people is always to consider it as informative in relation to some possible questions of this individual or group. We do not always realize this, because it is mostly assumed. It is assumed, for example, that a paper about the Sun may help answering questions about the Sun. It is less obvious, however, that a meteorite from outer space may answer questions about the origin of life. A good deal of scientific knowledge is needed to understand why this is the case (and a claim about the informativeness of something is knowledge-dependent and may turn out to be wrong). In a wider sense, background knowledge is always important for a person to extract information from any object (including documents and texts).

The Ontological Principle O1 well correlates with the assumption of Dretske (1981) that information is always relative to a receiver's background knowledge.

Some believe that dependence on prior knowledge in information extraction brings us to subjectivity in defining information and becomes the source of elusiveness of the concept of information (von Baeyer, 2004). Thus, the first impression is that the Ontological Principle O1 supports this subjective approach to the concept of information. However, in this case, subjectivity is confused with relativity. The Ontological Principle O1 states that information has to be considered not in the absolute way, as the majority of researchers in the field are doing, but as a relative essence properties of which depend on a chosen system. Dependence on an individual is usually called subjectivity. However, subjectivity is what depends only on the opinion of an individual. At the same time, information for a person *A* does not necessary coincides with what *A* thinks about information for herself or himself. For instance, *A* listens to a lecture and thinks that he gets a lot of information from it. This is a subjective estimate of information in the lecture. Nevertheless, if *A* forgets all he had heard the next day, the lecture actually has no information for *A*. This is an objective estimate of information in the lecture.

Another situation is when a person *B* reads some mathematical paper and finds nothing interesting there. Then *B* thinks that she has received no information from this paper. This is a subjective estimate of information in the paper. However, if *B* remembers something connected to this paper, then objectively she gets information from the text. Moreover, it is possible that *B* finds after some time that ideas from that paper are very useful for her work. This changes the subjective estimate of the paper and *B* starts to think that that paper contained a lot of useful information. Moreover, as axiological principles considered in (Burgin, 1995; 1997) show, for person *B*, information in that paper objectively also grows. This demonstrates that both objective and subjective estimates of information for a recipient depend not only on the recipient but also on time, interaction between the recipient and the carrier of information, work of this recipient and some other parameters.

Thus, information has objective but relativistic properties and their subjective estimates. This well correlates with the situation in the classical physics where objectivity is the pivotal principle.

The very fact that we treat science as a method of handling human experience inevitably involves the presence of observer in scientific theories (Lindsay, 1971). This allows us to better understand the role of the observer in interpretations of quantum theory. What seemingly began as a technical measurement problem in a specific area became gratuitously generalized into a metaphysical assertion that "observer-created" reality is all the reality that exists. The positivist idea that it is meaningless to discuss the existence of something which cannot be measured (position and velocity, within certain limits) has been developed into the idea that subatomic particles are unreal, formal structures, which only achieve actuality upon observation. In such a way, positivism became transformed into subjectivism (and even, solipsism), promoting the idea that the observer somehow creates reality by the act of observation. Heisenberg first stated that the electron does not have a well-defined position when it is not interacting. The next step in this direction is called the relational interpretation of quantum reality (Rovelli, 1996). It states that, even when interacting, the position of the electron is only determined in relation to a certain observer, or to a certain quantum reference system, or similar.

As Rovelli writes (1996), in physics, the move of deepening our insight into the physical world by relativizing notions previously used as absolute has been applied repeatedly and very successfully. The most popular examples are the Relativity Principle introduced by Galileo and relativity theory. By the Galileo's Principle of Relativity, the notion of the velocity of an object has been recognized as meaningless, unless it is indexed with a reference body with respect to which the object is moving. Thus, correct representation of motion demands a definite frame of reference. With special relativity, simultaneity of two distant events has been recognized as meaningless, unless referred to a specific light signal connecting these events.

In the light of the general theory of information, we can understand the relativity principle in quantum physics interpretation in the following way. Any (material) thing exists for people only when they get information from (in generalized sense, about) this thing. To get information from something, e.g., subatomic particle, we need an observer, i.e., recipient of information from this object. One may ask a question whether such particles existed before they were discovered. Einstein once asked Neils Bohr if the moon exists when no one is looking at it. Science gives a positive answer to such questions although, for example, there were no observers before these particles were discovered.

According to Heisenberg (1958), "The conception of objective reality of the elementary particles had thus evaporated … into the transparent clarity of a mathematics that represents no longer the behavior of particles but rather our knowledge of this behavior." According to the general theory of information, this knowledge is formed based on information that we get from particles.

The Ontological Principles O1 and O4 provide an information theoretical explanation for this. Indeed, for existence of something, it is necessary to consider observation in a generalized sense. Namely, we do not need any implication that the observer, or the observer system in quantum mechanics, is human or has any other peculiar property besides the possibility of interacting with the "observed" system *S*.

If we take the problem of subatomic particle existence for people before these particles were discovered, we see that recipients of information existed although those recipients did not know that they receive information from particles. For instance, psychological experiments show that people receive information but cannot identify it on the level of conscience (Luck, *et al*, 1996). Besides, there are several kinds of information in addition to cognitive information (Burgin, 2001). To explain this phenomenon and to solve the puzzle of physical existence, let us consider the following mental experiment. A particle, say electron, passes through a Wilson cloud chamber and produces a visible track of droplets condensed on ionized molecules. A digital or film camera makes a picture of this track. Only after 30 days, a physicist looks at this picture. It is evident that the electron existed before the observer looked at the picture and found evidence of its presence. Actually, we have information that the electron existed, at least, at the moment when it interacted with the molecules in the Wilson cloud chamber.

The Ontological Principle O1 also correlates with the idea of Roederer (2002) and some other researchers that interaction plays very important role in information processes. In other words, there exists no explicit information without interaction of the carrier of information with the receiver of information. However, it is possible to speak of information not only when we have both a sender and a recipient because the recipient can extract information from a carrier when the carrier does not send it. So, the classical *communication triad* (1) is not necessary for existence of information.

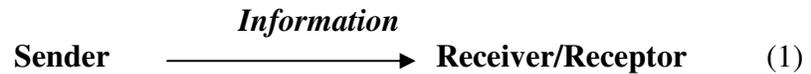
$$\textbf{Sender} \xrightarrow{\textit{Information}} \textbf{Receiver/Receptor} \quad (1)$$

The intrinsic necessary structure is the *input information triad* (2).

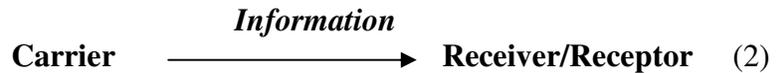
$$\textbf{Carrier} \xrightarrow{\textit{Information}} \textbf{Receiver/Receptor} \quad (2)$$

Note that in many situations, it is possible to treat a set or a sequence of carriers as one carrier. However, the structure of a carrier, e.g., whether it is integral or consists of separate parts, and the history of its interactions with the receptor can be important for some problems.

The triad (2) is complemented by the *output information triad* (3).

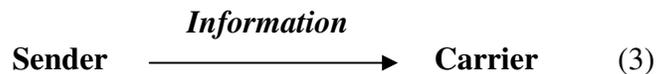
$$\textbf{Sender} \xrightarrow{\textit{Information}} \textbf{Carrier} \quad (3)$$

Together the output and input information triads form the communication triad (1) as their sequential composition. Note that it is possible that the carrier of information in the information triads (2) and/or (3) coincides with the Sender.

Besides, even if information gives some image of a pattern from a sender, this correspondence is not necessarily one-to-one.

It is also possible to speak about some implicit (potential) information in a carrier for a given system as a receptor.

Being more adequate to reality than previous assumptions about the essence of information, the first ontological principle makes it possible to resolve a controversy that exists in the research community of information scientists. Some suggest that information exists only in society, while others ascribe information to any phenomenon. Utilizing the Ontological Principle O1, general theory of information states that if we speak about information for people, then it exists only in society because now people exist only in society. However, when we consider a more general situation, then we see that information exists in everything and it is only a problem how to extract it.

Thus, the first principle explicates an important property if information, but says nothing what information is. This is done by the second principle that has several forms.

**Ontological Principle O2 (the General Transformation Principle).** *In a broad sense*, *information for a system **R** is a capacity to cause changes in the system **R***.

Thus, we can understand information in a broad sense as a capacity (ability or potency) of things, both material and abstract, to change other things.

This definition makes information an extremely widespread and comprehensive concept. Nevertheless, this situation well correlates with the etymological roots of the term *information*. This term originated from the Latin word '*informare*,' which can be translated as '*to give form to*,' '*to shape*,' or '*to form*.'

However, as it has happened with many other words, the meaning of the word information essentially changed. Since approximately the 16$^{th}$ century, the term *information* appears in ordinary French, English, Spanish and Italian in the sense we use it today: '*to instruct*,' '*to furnish with knowledge*', whereas the ontological meaning of '*giving form to something*' became more and more obsolete. Although, as Capurro thinks (1978; 1991), "*information* … came to be applied, as a more or less adequate metaphor, to every kind of process through which something is being changed or *in-formed*." This opinion strongly supports the Ontological Principle O2.

In addition, the Ontological Principle O2 well correlates with understanding of von Weizsäcker (2006/1985), who writes "we rate information by the effect it has" and with opinion of Boulding (1956), who writes that messages consist of information, while the meaning of a message is the change that it produces in the image. The idea that information is some change in the receiver was also proposed by MacKay (1956, 1961, 1969) although he restricted this change only to the cognitive system of the receiver.

Information is a general term. Like any general term, it has particular representatives. Such a representative is called a *portion of information*. For instance, information in this sentence is a portion of information. Information in this preprint is a portion of information. Information in a book is also a portion of information. Information in your head, dear reader, is also a portion of information.

To understand better the situation, let us consider some general terms. For instance, a book is a general term. The book that you are reading is a representative of this general term. A human being or a reader is a general term. At the same time, you, dear reader, is representative for both of these general terms.

Thus, the Ontological Principle O2 implies that information exists only in form of portions of information. Informally, a portion of information is such information that can be separated from other information.

**Remark 1.** In some cases, we use the term "information" instead of the term "a portion of information" when it does not cause misunderstanding.

We also consider such generic terms as a piece of information and a slice of information.

**Definition 2.** A *piece of information* is information that comes to a system in one interaction of this system.

Thus, a piece of information is also a portion of information. However, not any portion of information is a piece of information.

Note that the concept of a piece of information is relative, i.e., what is a piece of information for one system may be not a piece of information for another system.

**Definition 3.** A *slice of information* is a related to some object (domain, system or subject) portion of information.

MacKay suggested (1969) another quantization of information that takes into account the following two aspects:

1. The inner structure of the information element as a "logical a priori" aspect (structural information).

2. The so-called "weight of evidence" of the individual structural elements as an "empirical a posteriori" aspect (metrical information).

The unit of structural information is called *logon*. It is assumed that any portion of information can be divided into logons. Accordingly logon content, as a convenient term for the structural information-content, of a portion of information $I$ is defined as the number of logons in $I$.

The unit of metrical information is called *metron*. This unit, *metron*, is defined (MacKay, 1969) "as that which supplies one element for a pattern. Each element may be considered to represent one unit of evidence. Thus, the amount of metrical information in a pattern measures the weight of evidence to which it is equivalent."

MacKay (1969) tries to make this vague definition a little bit more exact, explaining that "the amount of metrical information in a single logon, or its metron-content, can be thought of as the number of elementary events which have been subsumed under one head or 'condensed' to form it."

As a result, the descriptive information is represented as an information vector in an information space.

Analyzing the Ontological Principle O2, we see that it has several consequences. First, it demonstrates that information is closely connected to transformation. Namely, it means that information and transformation are functionally similar because they both cause changes in a system. At the same time, they are different because information is a cause of change, while transformation is the change itself, or in other words, transformation is an operation, while information is what causes this operation.

Second, the Ontological Principle O2 explains *why* information influences society and individuals all the time, as well as why this influence grows with the development of society. Namely, reception of information by individuals and social groups induces transformation. In this sense, information is similar to energy. Moreover, according to the Ontological Principle O2, energy is a kind of information in a broad sense. This well correlates with the Carl Friedrich von Weizsäcker's idea (cf., for example, (Flükiger, 1995)) that *energy might in the end turn out to be information*. At the same time, the von Weizsäcker's conjecture explains the exact correspondence between a characteristic of thermal energy such as the thermodynamic entropy given by the Boltzmann-Planck formula $S = k \cdot \ln P$ and a characteristic of information such as the quantity of information given by a similar Hartley-Shannon formula $I = K \cdot \ln N$.

Third, the Ontological Principle O2 makes it possible to separate different kinds of information. For instance, people as well as any computer have many kinds of memory. It is even supposed that each part of the brain has several types of memory agencies that work in somewhat different ways, to suit particular purposes (Minsky, 1986). It is possible to consider each of these memory agencies as a separate system and to study differences between information that changes each type of memory. This would help to understand the interplay between stability and flexibility of mind, in general, and memory, in particular.

In addition, information that is considered in theory and practice is only cognitive information. At the same time, there are two other types: affective information and effective information (Burgin, 2001). For example affective information is very important for intelligence. As said Minsky (1998), "Emotion is only a different way to think. It may use some of the body functions, such as when we prepare to fight (the heart beats faster, etc.). Emotions have a survival value, so that we are able to behave efficiently in some situations. Therefore, truly intelligent computers will need to have emotions. This is not impossible or even difficult to achieve. Once we understand the relationship between thinking, emotion and memory, it will be easy to implement these functions into the software."

All this shows that the Ontological Principle O2 is very powerful. However, the common usage of the word information does not imply such wide generalizations as the Ontological Principle O2 does. Thus, we need a more restricted theoretical meaning because an adequate theory, whether of the information or of anything else, must be in significant accord with our common ways of thinking and talking about what the theory is about, else there is the danger that theory is not about what it purports to be about. Though, on the other hand, it is wrong to expect that any adequate and reasonably comprehensive theory will be congruent in every respect with common ways of thinking and speaking about its subject, just because those ways are not themselves usually consistent or even entirely clear. To achieve this goal, we use the concept of an *infological system* IF($R$) of the system $R$ to introduce information in the strict sense. It is done in two steps. At first, we make the concept of information relative and then we choose a specific class of infological systems to specify in the strict sense.

As a model example of an infological system IF($R$) of an intelligent system $R$, we take the system of knowledge of $R$. It is called in cybernetics the *thesaurus* Th($R$) of the system $R$.

Infological system plays the role of a free parameter in the general theory of information, providing for representation in this theory different kinds and types of information. Identifying an infological system IF($R$) of a system $R$, we can define information relative to this system. This definition is expressed by the following principle.

**Ontological Principle O2g (the Relativized Transformation Principle).** *Information for a system $R$ relative to the infological system* IF($R$) *is a capacity to cause changes in the system* IF($R$).

Now we can define information as the capacity of objects (things, texts, signals, etc.) to produce changes into infological system. In a more exact way, information for a system $R$ is the capacity of objects to produce changes into an infological system IF($R$) of $R$.

This definition is parallel to the definition of energy as the capacity of a physical system to do work, produce heat, light, electricity, motion, chemical reactions, etc. (Lindsay, 1971).

We can see that the direct result of the infological system concept introduction results in an even more general than in the Ontological Principle O2 definition of information. Indeed, in a general case, we can take as an infological system IF($R$) of the system $R$ any subsystem of $R$. In the case when IF($R$) of the system $R$, the Ontological Principle O2g becomes the same as the Ontological Principle O2. At the same time, taking different kinds of infological systems, it is possible to differentiate different kinds of information, while with respect to the Ontological Principle O2 all these kinds are not differentiated.

Generality of the Ontological Principle O2g allows one to consider energy as a specific kind of information. Namely, taking physical bodies (things) as a class **M** of infological systems, we see that information with respect to systems from **M** is what changes material bodies. However, we know that it is energy, which material bodies. Thus, energy is information that acts directly at material bodies. Other kinds of information act indirectly. Their action is meditated by energy.

However, being more general, such relativistic definition of information makes the concept more exact, flexible and selective than the concept of information in a broad sense introduced in the Ontological Principle O2. Information in a broad sense encompasses too much for the traditional understanding of information and intuition behind this that has been formed by practice of many generations.

The relativistic definition of information can be tuned up to the diversity of existing descriptions, interpretations, definitions, understandings and ideas of information. A choice of a definite infological system IF(*R*) in a system *R* allows a researcher, philosopher or practitioner to find such an interpretation of the concept of information that the best suits the goals, problems and tasks of this researcher, philosopher or practitioner.

The symbol IF may be considered as a name (denotation) of an operator that is defined in a space of systems of some chosen kind. It may be, in the space of all systems although the space of all system is a notion that, like the notion of all sets can cause contradictions (cf., for example, (Fraenkel and Bar-Hillel, 1958)). Being applied to a system *R*, the operator IF defines in *R* its infological system IF(*R*). For instance, taking such systems as people, we can correspond the mind to each individual *H* as an infological system IF(*H*) of this individual. Another option is correspond the system of knowledge of this individual *H* as her or his infological system IF(*H*). One more option is to correspond the system of beliefs (Bem, 1970) of this individual *H* as her or his infological system IF(*H*).

The infological system becomes a parameter of the definition and allows one to vary the scope, meaning and features of the defined concept. As a result, this parametric definition makes it possible to overcome the limitations in the theory of information imposed by the, so-called, Capurro trilemma (Fleissner and Hofkirchner, 1995; Capurro, Fleissner and Hofkirchner, 1999). This trilemma states that information may mean either the same at all levels (*univocity*) or something similar at all levels (*analogy*) or something different at different levels (*equivocity*). In the first case, as Capurro suggests (Capurro, Fleissner and Hofkirchner, 1999), "we lose all qualitative differences, as for instance, when we say that e-mail and cell reproduction are the same kind of information process. Not only the "stuff" and the structure but also the processes in cells and computer devices are rather different from each other. If we say the concept of information is being analogically used, then we have to state what the "original" meaning is." If the concept of information is considered at the human level, then we are confronted with anthropomorphism when we use the same concept at a non-human level. By the same token, we would say that "in some way" atoms "talk" to each other, etc. Finally there is equivocity, which means that, for example, information in physics and information in education are wholly different concepts. In this case, information cannot be a unifying concept any more. Further reasoning bring Capurro to the conclusion that "we are faced with infinite concepts of information, something which cannot be overlooked by any kind of theory."

However, having a sufficiently variable parameter, we can generate a family of concepts that represent existing understandings and interpretations of the word *information*. The relativistic definition of information provides such a flexible parameter as the infological system. This definition possesses, at the same time, univocity, analogy and equivocity. As a result, this definition encompasses, at the same time, the broad concept of information given in the Ontological Principle O2 and a variety of more restricted concepts. Examples of such more restricted concepts are: information as *elimination of uncertainty* (statistical approach), *making a distinction* (the approach suggested by Hofkirchner (1999)) or "*any difference that makes a difference*" (the approach suggested by Bateson (2000)).

Any system can have an infological system and not a single one. Consequently, in contrast to the opinion of some researchers, information is important both for the biotic and abiotic worlds. Information enters non-living physical world even without living beings.

This implies that for a complex system there are different kinds of information. Each type of the infological system determines a specific kind of information. For example, information that causes changes in the system of knowledge is called cognitive information. Existing approaches to information theory and problems with understanding information as natural, social, and technological phenomenon resulted in a current situation when researchers consider only cognitive information.

At the same time, Roederer defines information as *the agent that mediates the correspondence between features or patterns in the source system A and changes in the structure of the recipient B*. This definition strongly correlates with the definition from the Ontological Principle O2a. Taking such infological system as genetic memory, we come to the concept of biomolecular information considered by Roederer (2002).

The concept of infological system is very flexible. Indeed, it is possible to consider even dynamic infological systems. As Scarrott writes (1989), the most basic function of information is to control action in an organized system and thereby operate the organization. The concept of infological system allows one to reflect this situation, taking behavior of a system as a dynamic infological system.

Information is such a profound essence that to be able to discern new types of information and to become more specific about information *per se* or information in strict sense, we need better understanding of the structure of the world where we live.

As we know, people live in the physical (material) world and many perceive that this is the only reality that exists. However, some Eastern philosophical and religious systems, e.g., Buddhism, teach that physical reality is a great illusion and the only reality is the spiritual world. As science does not have enough evidence to accept or reject this idea, we are not going to discuss it. Nevertheless, science has enough evidence to accept existence of the mental world. As states contemporary psychology, each individual has a specific inner world, which is based on the psyche and forms mentality of the individual. These individual inner worlds form the lowest level of the mental world, which complements our physical world.

Some thinkers, following Descartes, consider the mental world as independent of the physical world. Others assume that mentality is completely generated by physical systems of the organism, such as the nervous system and brain as its part. However, in any case, the mental world is different from the physical world and constitutes an important part of our reality.

Moreover, our mentality influences the physical world and can change it. We can see how ideas change our planet, create many new things and destroy existing ones. Even physicists, who research the very foundation of the physical world, developed the, so-called, observer-created reality interpretation of quantum phenomena. A prominent physicist, Wheeler, suggests that in such a way it is possible to change even the past. He stresses (Wheeler, 1977) that elementary phenomena are unreal until observed. This gives a dualistic model of reality.

However, the dualistic model is not complete. This incompleteness was prophesized in ancient Greece and proved by modern science. One of the great ideas of ancient Greece is the world of ideas (or forms), the existence of which was postulated by Plato. In spite of the attractive character of this idea, the majority of scientists and philosophers believe that the world of ideas does not exist, because nobody has any positive evidence in support of it. The crucial argument of physicists is that the main methods of verification in modern science are observations and experiments, and nobody has been able to find this world by means of observations and experiments. Nevertheless, there are modern thinkers who, like such outstanding scholars as philosopher Karl Popper, mathematician Kurt Gödel, and physicist Roger Penrose, continue to believe in the world of ideas, giving different interpretations of this world but suggesting no ways for their experimental validation.

However, science is developing, and this development provided recently for the discovery of the world of structures. On the level of ideas, this world may be associated with the Platonic world of ideas in the same way as atoms of modern physics may be related to the atoms of Democritus. The existence of the world of structures is demonstrated by means of observations and experiments. This world of structures constitutes the structural level of the world as whole. Each system, phenomenon, or process either in nature or in society has some structure. These structures exist like things, such as tables, chairs, or buildings, and form the structural level of the world. When it is necessary to investigate or to create some system or process, it is possible to do this only by means of knowledge of the corresponding structure. Structures determine the essence of things.

Let consider the development of ideas related to the *global world structure*.

In the Platonic tradition, the global world structure has the form of three interconnected worlds: *material*, *mental*, and the *world of ideas* or *forms*.

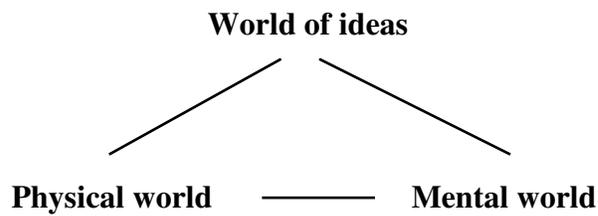

**Figure 1**. The Plato triad of the world

However, existence of the world of ideas has been severely criticized. Many argue that taking a long hard look at what the Platonist is asking people to believe, it is necessary to have faith in another "world" stocked with something called ideas. This results in many problems. Where is this world and how do we make contact with it? How is it possible for our mind to have an interaction with the Platonic realm so that our brain state is altered by that experience? Plato and his followers have not provided convincing answers to these questions.

Popper's ontology consists of three worlds:

**World 1**: Physical objects or states.

**World 2**: Consciousness or psychical states.

**World 3**: Intellectual contents of books, documents, scientific theories, etc.

As Popper uses the words *information* and *knowledge* interchangeably, World 3 consists of knowledge and information and we have the following triad.

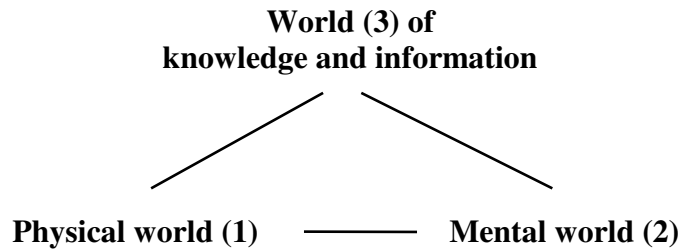

**Figure 2.** The Popper triad of the world

The Popper triad is much more understandable than the Plato triad because people know what knowledge is much better than what ideas are, especially, when these are Plato ideas, or forms.

Other authors refer World 3 to signs in the sense of Charles Pierce, although they do not insists that World 3 consists of objects that Pierce would classify as signs (cf., for example, (Skagestad, 1993; Capuro and Hjorland, 2003)).

Only recently, modern science made it possible to achieve a new understanding of Plato ideas, representing the global world structure as the *Existential Triad* of the world. In this triad, the Physical (material) World is interpreted as the physical reality studied by natural sciences, while ideas or forms might be associated with structures, and the Mental World encompasses much more than individual conscience (Burgin, 1997; Burgin and Milov, 1999). In particular, the Mental World includes social conscience. This social conscience is projected on the collective unconscious in the sense Jung (cf. (Jung, 1969)) by the process of internalization (Atkinson, et al, 1990). In addition, the World of structures includes Popper's World 3 as knowledge, or the intellectual contents of books, documents, scientific theories, etc., is a kind of structures that are represented in people's mentality (Burgin, 1997; 2004).

Thus, the *existential triad* of the world (the world's global structure) has the following form:

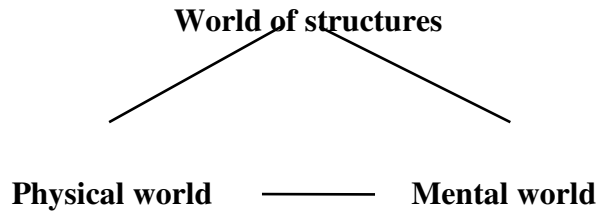

**Figure 3.** The Existential Triad of the world

In the mental world, there are real "things" and "phenomena". For example, there exist happiness and pain, smell and color, love and understanding, impressions and images (of stars, tables, chairs and etc.). In the physical world, there are the real tables and chairs, sun, stars, stones, flowers, butterflies, space and time, molecules and atoms, electrons and photons. It has been demonstrated (Burgin, 1997) that the world of structures also exists in reality. For instance, the fundamental triad described in (Burgin, 2004a) exists in the same way as tables, chairs, trees, and mountains exist. Knowledge, *per se*, forms a component of the world of structures. It is an important peculiarity of the world (as a whole) that it exists in such a triadic form not as a static entity but as a dynamic structure.

It is necessary to understand that these three worlds are not separate realities: they interact and intersect. Thus, individual mentality is based on the brain, which is a material thing. On the other hand, physicists discuss a possibility that mentality influences physical world (cf., for example, (Herbert, 1987)), while our knowledge of the physical world to a great extent depends on interaction between mental and material worlds (cf., for example, (von Baeyer, 2001)).

Even closer ties exist between structural and material worlds. Actually no material thing exists without structure. Even chaos has its chaotic structure. Structures do things what they are. For instance, it is possible to make a table from different material: wood, plastics, iron, aluminum, etc. What all these things have in common is not their material; it is specific peculiarities of their structure. As argue some physicists, physics studies not physical systems as they are but structures of these systems, or physical structures. In some sciences, such as chemistry, and areas of practical activity, such as engineering, structures play a leading role. For instance, the spatial structure of atoms, chemical elements, and molecules determines many properties of these chemical systems. In engineering, structures and structural analysis even form a separate subject (cf., for example, (Martin, 1999)).

We can see existential triads in every individual and in each computer. An individual has the physical component - her body, the mental component studied by psychologists, and the structural component, which comprises all structures and systems from the other two components. A computer has the physical component - its hardware, the mental component, which consists of everything that is in computer memory, and a structural component, which comprises all structures and systems from the other two components.

While physical and mental objects have been treated for a long time and thus, do not cause difficulties in understanding, structure is a much more imprecise concept and needs additional explanation. Here we do not give formal definitions of concepts related to the World of Structures as the main topic of the book is information.

**Definition 2.1.4.** A *structural representation* R(**K**) of an entity (system, process) **K** is an abstract (symbolic or mental) image (representation) of **K** consisting of representations of parts (elements, components) of **K** and connections (ties, relations) between them.

Each system has inherent structural representations. At the same time, inherent structural representations are (better or worse) revealed by reflected structural representations situated in other systems.

There are three main types of structural representations:

1. *Material* in the Physical World.
2. *Symbolic* in the Mental World.
3. *Intrinsic* in the Structural World.

**Definition 4.** A *structure* S(**K**) is a class of equivalent structural representations $R_i$(**K**).

This equivalence depends on the level of structuring. This level determines what elements and ties are distinguished. For instance, let us consider three functions: *f* from the set *N* of all natural numbers into itself, *g* from the set *R* of all real numbers into itself, and *h* from the set *N* into the set *R*. In mathematics, these functions are denoted by *f*: *N* → *N*, *g*: *R* → *R*, and *h*: *N* → *R*. If we treat sets *N* and *R* as members of the class with the name *a mathematical object* and do not discern them, then all three functions have the same structure of the following fundamental triad

**mathematical object 1  →  mathematical object 2**

In contrast to this, if we look upon sets *N* and *R* as different mathematical object, then all three functions have the different structures, although all of them are isomorphic to the fundamental triad. In the first case, we do not pay attention at distinctions between sets *N* and *R*, while in the second case, such distinctions are essential. As a result, the same systems (objects) acquire different structures: in the first case, the structure is ●→●, while in the second case, the structure is ●→○.

It is possible to ask whether object representation allows us to get objective structures of objects, i.e., to explicate structures that exist in reality, or the structure of a system depends only on those who build the representation. At first glance, the process of structure explication looks very subjective as it rests on how distinguishability of elements and ties is defined. However, similar situation exists with what people see, hear and describe. For instance, an individual sees a table. In this situation, we can ask the question, *Is it an image of a real table, an image on a screen or only an illusion of the brain*?

Some philosophers, e.g., Berkeley, as well as some religious systems, e.g., Buddhism, argue that everything what people perceive by their senses is an illusion.

In contrast to this, science developed means to discern objective reality from illusions. These means are based on repetition of observations and experiments. Although these methods are not perfect, they provide for the whole existence of science and development of the human civilization based on science as a source of the technological progress.

Analyzing situation with structures as invariant representations of existing things (real or abstract systems), we come to the conclusion that structures exist in the same way as physical things, such as tables, chairs, and buildings, exist. We do not give here explicit proofs of this deep-rooted statement. It is possible to find details of these proofs in (Burgin, 1997).

Formalized abstract structures are studied by pure mathematics. Formalized and informal concrete structures are studied by all sciences (natural, social and life sciences), applied mathematics and philosophy.

In modern science and philosophy a general approach based on revealing intrinsic structures kept invariant relative to transformations is called "structuralism". Structuralists' point of view is that different things and phenomena have identical essence if their structures are identical. To investigate a phenomenon means for them to find its structure.

From the beginning structuralism have appeared in linguistics. Although F. de Saussure did not use the term "structure" but the term "form", his understanding in many respects coincides with the structuralistic approach. The terms "structure" and " the structural linguistics" was introduced by V.Brendal (1939). L. Hjelmslev (1958) considered structure as "independent essence with inner dependencies". He emphasized that the properties of each element of such integrity depend on its structure.

Afterwards, structuralism was extended to other fields. Levi-Strauss developed and successfully applied the structuralistic methods and structural analysis in anthropology. On this basis he had restored the common (for all cultures) function which intermediates the fundamental contradictions of the human existence. This function (in his opinion) has been lost by the modern European civilization.

Lacan (1977) applied the structural analysis to psychoanalysis. Subconscious (in his opinion) is structured as a language. It is possible to describe his main conception by the triad:

**real - imaginary - symbolic**

Real is treated as chaos, inaccessible is a name. Imaginary is an individual variation of the symbolical order. Symbolical order is objective. Lacan has been proving the thesis that an idea and existence are not identical: it is a language, which is a mediator between them.

The transition to the study of a structural reality reflects a new stage of the structuralistic doctrine. In particular, investigation of the structural level of the world and scientific research of structures as real phenomena has made possible to find the most basic structure called *fundamental triad* or *named set*.

In a symbolic form, a *named set* (*fundamental triad*) **X** is a triad $(X, f, I)$ where $X$ is the *support* of **X** and is denoted by S(**X**), $I$ is the *component of names* (also called *set of names* or *reflector*) of **X** and is denoted by N(**X**), and $f$ is the *naming correspondence* (also called *reflection*) of the named set **X** and is denoted by n(**X**). The most popular type of named sets is a named set **X** = $(X, f, I)$ in which $X$ and $I$ are sets and $f$ consists of connections between their elements. When these connections are set theoretical, i.e., each connection is represented by a pair $(x, a)$ where $x$ is an element from $X$ and $a$ is its name from $I$, we have a *set theoretical named set* or binary relation. Bourbaki in their fundamental monograph (1960) also represent binary relations in a form of a triad (named set).

However, the study of the World of Structures is only at its beginning.

The structure of the world induces similar structures in many systems that exist in this world. It is true for language, symbols, social institutions and organizations, psyche, intelligence and many others. In particular, the world structure has definite implications for infological systems.

When ***R*** is a material system, its infological system IF(***R***) consists of three components:

- the *material component*, which is a system of physical objects;

- the *symbolic component* realized by the material component;
- the *system of structures*, which form the structural component of the infological system.

The symbolic component plays the role of a connecting link between the material and structural components. To understand this, let us consider theoretical understanding embodied in models for the concept of a symbol.

If we analyze the usage of the word "symbol," we come to the conclusion that it has three different, however, connected, meanings. In a broad sense, symbol is the same as sign. For example, the terms "symbolic system" and "sign system" are considered as synonyms, although the first term is used much more often. The basic property of the sign is that sign points to something different than itself, transcendent to it.

The second understanding of the word "symbol" identifies symbol with a physical sign, that is, some elementary entity inscribed on paper, papyrus or stone, presented on the screen of a computer monitor, and so on. Letters are signs and symbols at the same time. Decimal digits 0, 1, 2, 3, 4, 5, 6, 7, 8, and 9 are also signs and symbols.

However, we are interested in the third meaning of the word "symbol" when it is considered in a strict sense. Such understanding was developed in *semiotics* as a general theory of signs. Semiotics studies structure of signs and their communicative function. As signs exist in a huge diversity of situations, the founder of semiotics, Charles Pierce, and his follower Charles Morris, defined semiotics very broadly, in the hope that it would influence as many disciplines as possible. For instance, Morris wrote (1938):

"The sciences must look to semiotic for the concepts and general principles relevant to their own problems of sign analysis. Semiotic is not merely a science among other sciences but an organon or instrument to all sciences."

Indeed, today semiotics is an important tool in communication research, information theory, linguistics and the fine arts, as well as in psychology, sociology and esthetics. Yet, although many other disciplines recognize the potential importance of semiotic paradigms for their fields, they have not yet found a satisfying way of integrating them in their domain.

While many use the word *symbol* in the same contexts as the word *sign*, the French linguist Ferdinand de Saussure, sometimes called the father of theoretical linguistics, understood "sign" as a category under "symbol" (Saussure, 1916). To represent the main property of signs, de Saussure introduced a structural model of sign in the form of the *dyadic sign triad* by (see figure 4).

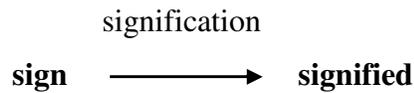

**sign** → **signified**

(signification above the arrow)

**Figure 4.** The dyadic sign triad of de Saussure

This triad is a kind of the fundamental triad.

Considering the relation between the concepts *sign* and *symbol*, Pierce inverted this relation, making *sign* the general term and *symbol* its particular case as the convention-based sign. According to Pierce, there are three kinds of signs: *icon*, *index*, and *symbol*.

The dyadic sign triad explicates important properties of sign, but not all of them. Namely, sign represents something different than itself due to the meaning. That is why Pierce extended this dyadic model by further splitting the signified into essentially different parts: the sign's object and interpretant (meaning of the sign), and thus, coming to the triadic model of a sign, the *balanced sign triad*:

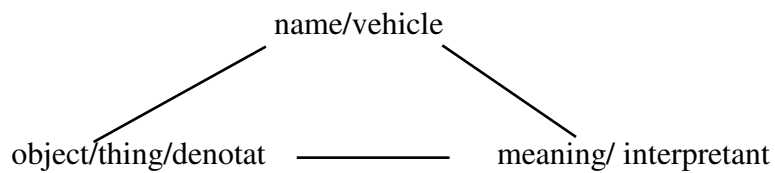

**Figure 5.** The balanced sign triad of Pierce

Thus, in the model of Pierce, a sign is understood as a relation consisting of three elements: Name (Vehicle), Object and Meaning of the sign. Usually, a physical representation of a sign is what people call sign in everyday life, that is, a sign is some elementary image inscribed on paper, clay tablet, piece of wood or stone, presented on the screen of a computer monitor, and so on.

The balanced sign triad is similar to the existential triad of the World. In it, the component *name* corresponds to the structural world as a syntactic system; the component *object/denotat* correspond (but does not necessarily belongs) to the physical world; and the component *meaning/interpretant* corresponds to the mental world as a semantic system. In many cases, the object is a material thing and as such, is a part of the physical world. However, object can be non-material and thus, does not belong to the physical world in some cases. For instance, the word *joy* is the name of emotion, which is nonmaterial. The word *theorem* is the name of a mathematical statement with definite properties. A statement is also nonmaterial. Nevertheless, object as a component of sign plays the same role as thing, implying that the Pierce triad is homomorphic to the existential triad.

An *icon* looks like what it signifies. Photographs at the level of direct resemblance or likeness are therefore heavily iconic. We all are familiar with computer icons, that helped popularize such a word processor as the Word, as well as with the pictographs such as are used on "pedestrian crossing" signs. There is no real connection between an object and an icon of it other than the likeness, so the mind is required to see the similarity and associate an object and its icon. A characteristic of the icon is that by observing it, we can derive information about the object the icon signifies. The more simplified the image, the less it is possible to learn. No other kind of signs gives that kind of pictorial information.

Pierce divides icons further into three kinds: images, icons and symbols. *Images* have the simplest quality, the similarity of aspect. Portraits and computer icons are images. *Diagrams* represent relationships of parts rather than tangible features. Examples of diagrams are algebraic formulae. Finally, *metaphors* possess a similarity of character, representing an object by using a parallelism in some other object. Metaphors are widely used in poetry and language, for example, the frogs of Aesop who desired a king, computer mouse or vector fields in physics.

An *index* has a causal and/or sequential relationship to its signified. A key to understanding indices (or indexes) is the verb "indicate", of which "index" is a substantive. For instance, indices are directly perceivable events that can act as a reference to events that are not directly perceivable, or in other words, they are something visible that indicates something out of sight. You may not see a fire, but you do see the smoke and that indicates to you that a fire is burning. Words "this", "that", "these", and "those" like a pointed finger, are also indices. The nature of the index can be unrelated to that of the signified, but the connection here is logical and organic - the two elements are inseparable - and there is little or no participation of the mind.

Contemporary microphysics is built on indices called subatomic, or subnuclear, particles. Physicists cannot see these particles even with the best modern microscopes. However, physicists see the results of particle interactions, e.g., in the Wilson cloud chamber, and know properties of these particles.

A *symbol* represents something in a completely arbitrary relationship. The connection between signifier and signified depends entirely on the observer, or more exactly, what the observer was taught or invented. Symbols are subjective. Their relation to the signified object is dictated either by social and cultural conventions or by habit. Words are a prime example of signs. Whether as a group of sounds or a group of characters, they are only linked to their signified because we decide they are and because the connection is neither physical nor logical, words change meaning or objects change names as time goes by. Here it all happens in the mind and depends on it.

However, often, especially in science, people try to create words so that they show/explicate connections to the signified. For instance, a computer is called *computer* because it/he/she computes. A teacher is called *teacher* because she/he teaches. Some elementary particles are called neutrons because they are electrically neutral, i.e., their electrical charge is zero.

Symbols are abstract entities, and whenever we use one, we are only pointing to the idea behind that symbol. Do you know how computer aliases (or shortcuts) work? You create a file that opens the actual file it refers to. If you trash the alias/shortcut, it does not affect the file. Symbols work in exactly the same way in relation to the concept they serve. The @, © and $ symbols, logical symbols, astrological symbols, road signs, V of victory, all are symbols.

Pierce divides symbols further into two kinds: a *singular symbol* denotes tangible things, while an *abstract symbol* signifies abstract notions. However, it is not always easy to make a distinction. For example, such symbol as "lion" signifies an abstract notion of a lion as a specific animal. At the same time, this symbol as "a lion" signifies the set of all lions. Thus, it is more tangible to introduce one more class of symbols, which we call general symbols. A *general symbol* signifies both an abstract notion and a collection of things encompassed by this notion. For example, "a lover" is a general symbol, while "love" is an abstract symbol.

One and the same word can be used as a name for different symbols and even for different types of symbols. For instance, on the social level, the word a "field" is used as an individual symbol when it denotes a specific place on the Earth. At the same time, it will be an abstract symbol used in mathematical community and denoting a specific mathematical structure, or more exactly, two kinds of structures – fields in algebra, such as the field of all real numbers, and fields in functional analysis, such as a vector field. On another, wider group level, the same word is used as a name of some system, such as a field of mathematics, field of activity or field of competence. Important examples of symbols are general concepts and formal expressions.

For example, the material component of the infological system of a human being is the brain or its part that is called memory. What is commonly called memory is not a single, simple system. It is an extraordinarily complex system of diverse components and processes. Memory of a person has three, or perhaps even more, distinct components (Minsky, 1986). The most important and best documented by scientific research are sensory information storage (SIS), short-term memory (STM), and long-term memory (LTM). Memory researchers do not employ uniform terminology. Sensory information storage is also known as sensory register, sensory store, and eidetic and echoic memory. Short- and long-term memories are also referred to as primary and secondary memory. Each component of memory differs with respect to function, the form of information held, the length of time information is retained, and the amount of information-handling capacity. Memory researchers also posit the existence of an interpretive mechanism, as well as an overall memory monitor or control mechanism that guides interaction among various elements of the memory system.

The corresponding to the brain symbolic component is the mind. In the case of the memory as the material component, we take the symbolic representation of the data and knowledge system as the corresponding symbolic component.

Actually any entity can be used as a symbol, even a process. For instance, neurons do not have memory. However, neural networks allow one to organize some kind of memory, being capable of storing and retrieving data from this memory. There are two formats for saving data: static and dynamic. Dynamic storage is utilized for temporary data. In this case, a part of a network is used only for preserving information. If we take a neuron, which can be in two states: firing and silent, then it is possible to interpret a silent neuron as containing the symbol '0', while a firing neuron is considered as containing the symbol '1'. When a neuron can fire several kinds of output, for example any rational number, then we can store in it more than two symbols. To preserve the firing state, a neuron can be initiated to circulate in a loop, until it is stopped. Since the output of the neuron feeds back to itself, there is a self-sustaining loop that keeps the neuron firing even when the top input is no longer active. Activating the lower input suppresses the looped input, and the node stops firing. The stored binary bit is continuously accessible by looking at the output. This configuration is called a latch. Thus, a symbol, e.g., 1, is stored in form of a process in a neural network. Consequently, this process becomes a symbol itself. Existence of such memory is supported by the experimental evidence that some patterns in the brain are preserved in a dynamical fashion (Suppes and Han, 2000).

The corresponding to the brain structural component includes knowledge of the individual. Structural component also includes beliefs, attitudes (Bem, 1970), images, ideas, conjectures, problems, etc. However, we can take the system of knowledge, also called thesaurus, as an infological system of the brain, or of an individual. Infological elements in this case will be units of knowledge of the individual.

Another example of an infological system is the memory of a computer. Such a memory is a place in which data and programs are stored. Data and programs are represented by symbols in a material form: as states of electronic elements or written on paper, board or some other material objects. At the same time, data texts, symbols, knowledge, programs, algorithms and many other essences are structures (Burgin, 1997; 2004; 2005).

The computer memory is also a complex system of diverse components and processes. Memory of a computer includes such three components as the random access memory (RAM), read-only memory (ROM), and secondary storage. While RAM forgets everything whenever the computer is turned off and ROM cannot learn anything new, secondary storage devices allow the computer to record information for as long period of time as we want and change it whenever we want. Now the following devices are utilized for log-term computer memory: magnetic tapes and corresponding drives, magnetic disks and corresponding drives, and optical disks and corresponding drives.

**Remark 2.** In an arbitrary system *R*, it is possible to select different infological systems. Fixing one of these subsystems, we determine the type of information for *R* and changing our choice of IF(*R*) we change the scope of information entities (portions of information).

For example, computers have different kinds of memory: processor registers, addressed memory, main storage, buffer storage, external storage, working storage etc. Each of them or any combination of them may be considered as an infological system of a computer *R*. If the processor registers are treated as an infological system, then a program (even such that is kept in the main storage of this computer) does not have information for *R* until execution of instructions from this program begins.

**Definition 4.** Elements from IF(*R*) are called *infological elements*.

There is no exact definition of infological elements although there are various entities that are naturally considered as infological elements as they allow one to build theories of information that inherit conventional meanings of the word *information*. For instance, knowledge, data, images, ideas, fancies, abstractions, beliefs, and similar objects are standard examples of infological elements. If we consider only knowledge and data, then the infological system is the *system of knowledge* of a given system ***R***. Such a system of knowledge is called a *thesaurus* in cybernetics.

The situation with infological elements looks similar to the situation in contemporary physics where physicists do not give a definition of matter but explain how matter is built and what elements of matter are. As such elements on the lowest level of the hierarchy, physicists take subatomic particles, physical fields, atoms, molecules and so on.

The infological system plays the role of a free parameter in the definition of information (cf. the Ontological Principle O2g). One of the ways to vary this parameter is to choose a definite kind of infological elements. Additional conditions on infological elements imply a more restricted concept of information.

To better understand how infological system can help to explicate the concept of information in the strict sense, we consider cognitive infological systems.

**Definition 5.** An infological system IF(***R***) of the system ***R*** is called *cognitive* if IF(***R***) contains (stores) such elements or constituents as knowledge, data, images, ideas, fancies, abstractions, beliefs, etc.

Cognitive infological system is a standard example of infological systems, while its elements, such as knowledge, data, images, ideas, fantasies, abstractions, and beliefs, are standard example of infological elements. Cognitive infological system is very important, especially, for intelligent systems. The majority of researchers believe that information is intrinsically connected to knowledge (cf. Flückiger, 1995).

The system of knowledge KIF(***R***) of ***R*** is an infological system of an intelligent system ***R***. In cybernetics, it is called the thesaurus Th(***R***) of the system ***R***.

A thesaurus is a part of a cognitive infological system. Another example of an infological system is the memory of a computer. Such a memory is a place in which data and programs are stored.

A cognitive infological system of $R$ is denoted by CIF($R$) and is related to cognitive information.

**Ontological Principle O2c (the Cognitive Transformation Principle).** *Cognitive information for a system $R$, is a capacity to cause changes in the cognitive infological system* IFC($R$) *of the system $R$.*

As the cognitive infological system contains knowledge of the system it belongs, cognitive information is the source of knowledge changes. This perfectly correlates with the approach of Dretske (1983) and Goldman (1967) who defined knowledge as information-caused belief, i.e., information produces beliefs that are called knowledge. Moreover, it is impossible to obtain knowledge without information. Dretske (1983) develops this idea, implying that information produces beliefs, which, according to our definition, are also elements of the cognitive infological system. Moreover, many researchers relate all information exclusively to knowledge. For instance, Mackay writes (1969):

"Suppose we begin by asking ourselves what we mean by information. Roughly speaking, we say that we have gained information when we know something now that we didn't know before; when 'what we know' has changed."

Barwise and Seligman write (1997) that "information is closely tied to knowledge." Cognitive information related to knowledge was also studied by Shreider (1967).

At the same time, other researchers connect cognitive information to experience. For instance, Boulding (1956) calls a collection of experiences by the name *image* and explains that messages consist of information as they are structured experiences, while the meaning of a message is the change that it produces in the image.

Cognitive information is what people, as a rule, understand and mean when they speak about information. Indeed, since approximately the $16^{th}$ century, we find the word *information* in ordinary French, English, Spanish and Italian in the sense we use it today: to instruct, to furnish with knowledge (Capurro, 1991). However, scientific usage of the notion of information (cf, for example, (Loewenstein, 1999)) implies a necessity to have a more general definition. For instance, Heidegger pointed to the naturalization of the concept of information in biology in the form of genetic information (Heidegger and Fink, 1970). An example, of such a situation is when biologists discuss information in DNA in general or in the human genom, in particular.

As a result, we come to the world of structures. Change of structural features of a system, and through them other system characteristics, is the essence of information in the strict sense. This correlates with the von Weizsäker's remark that information being neither matter nor energy, according to (Wiener, 1961), has a similar status as the "platonic eidos" and the "Aristotelian form" (Weizsäcker, 1974).

**Ontological Principle O2a (the Special Transformation Principle).** *Information in the strict sense or*, *simply*, *information for a system **R***, *is a capacity to change structural infological elements from an infological system* IF(***R***) *of the system **R***.

To understand this principle and the definition of information it contains, we need to understand the concept of a structure. Otherwise, this definition will be incomplete, containing an undefined term. This brings us to the most fundamental question of ontology how our world is organized (built).

Some researchers related information to structure of an object. For instance, information is characterized as a property of how entities are organized and arranged, not the property of entities themselves (Reading, 2006). Other researchers related information to form and form is an explicit structure of an object. For instance, information is characterized as an attribute of the form (in-*form*-ation) that matter and energy, not of the matter and energy themselves (Dretske, 2000).

However, absence of the exact concept of structure and lack of understanding that structures can objectively exist result in contradictions and misconceptions. For instance, a researcher writes that information is simply a construct used to explain causal interaction, and in the next sentence, the same researcher asserts that information is a fundamental source of change in the natural world. However, constructs cannot be sources of change, they can only explain change.

The Ontological Principle O2a implies that information is not of the same kind as knowledge and data, which are structures (Burgin, 1997). Actually, if we take that *matter* is the name for all substances as opposed to *energy* and the *vacuum*, we have the relation that is represented by the following diagram called the Structure-Information-Matter-Energy (SIME) Square.

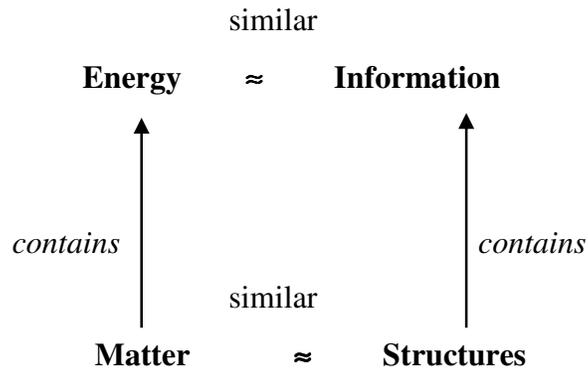

**Figure 6.** The Structure-Information-Matter-Energy (SIME) Square

In other words,

***Information is related to knowledge and data as energy is related to matter.***

Here it is necessary to remark that many people think and write that there is no distinction between matter and energy. They base their belief on the famous formula from the relativistic physics

$$E = mc^2 \qquad (4)$$

However, this belief is only a misconception because the letter *m* in (4) does not stand for matter, as well as the letter *E* in (4) does not stand for energy. Here *m* means "mass" and *E* means the quantity of energy. "Mass" is only one of the characteristics or measures of material objects. What concerns the symbol *E* in (4), it is necessary to make a distinction between energy as a natural phenomenon and energy as some physical quantity. The latter is a measure of the former. It is natural to cal this quantity by the name "the quantity of energy." However, traditionally it is also called energy. There are other measures of energy. For instance, entropy is another measure of thermal energy.

Besides, what Einstein really proved was that if a body at rest emits a total energy of *E* remaining at rest, then the mass of this body decreases by $E/c^2$.

Thus, formula (4) gives a relation between two measures of two distinct natural phenomena. Namely, this formula estimates how much energy is stored in matter.

The reasoning that formula (4) means absence of distinction between energy and matter is similar to the following argumentation. Let M be a man, T be a tree, and *h(x)* denotes the height of *x*. Then some can (incorrectly) say that the formula *h*(M) = *h*(T) means that there is no distinction between M and T. Although this fallacy is more evident than the fallacy of equating energy and matter, both fallacies have the same nature.

It is important to understand that saying or writing that matter contains energy or knowledge contains information is not the same as saying that a bottle contains water. The meaning of the expression "knowledge contains information" is similar to the meaning of expressions "the brain contains knowledge" or "a person has knowledge." In other words, it is possible to extract energy from matter, as well as it is possible to extract information from knowledge. In some cases, such extraction goes on automatically (on the unconscious level). It gives an illusion that information comes itself into a system.

It is possible to reproach that the concept of an infological system is too ambiguous and fuzzy. However, ambiguity may be a positive property if you can use it. For example, if you can control and change ambiguity, it becomes not an ambiguity but a parameter that is utilized to tune and control the system.

This is just the case with the infological system in general theory of information. Thus, it is natural that considering a human being, we do not chose the same infological systems as we do for biological cells or computers. Besides, any complex system has, as a rule, several infological systems.

The main infological system of an individual is the mind with its material component – the brain. This infological system controls all human actions. As a result, behavior of people is determined by information that flows in the organism of this individual. It allows individual to adapt to environment both in nature and society.

A possibility to choose an infological system in a different way is very beneficial. It explicates existence of different types and kinds of information. Each type of information corresponds to some type of infological system. Examples of such types are considered elsewhere.

In what follows, we consider only information in the strict sense.

It is possible to separate (cf., for example, (Flükiger, 1995)) three approaches to information: information as a thing, information as a structure, and information as a property. It is possible to ask which of these approaches is true. The general theory of information supports the third approach.

Let us consider a situation when one system *Q* sends a signal to another system *R*. This signal carries a message, which, in turn, contains information.

Let *I* be a portion of information for a system *R*.

**Ontological Principle O3 (the Embodiment Principle).** *For any portion of information I, there is always a carrier C of this portion of information for a system R.*

Really, people get information from books, magazines, TV and radio sets, computers, and from other people. To store information people use their brains, paper, tapes, and computer disks. All these entities are carriers of information.

For adherents of the materialistic approach, the Ontological Principle O3 must be changed to its stronger version.

**Ontological Principle OM3 (the Material Embodiment Principle).** *For any portion of information I, there is some substance C that contains I.*

For example, according to Landauer (2002), information is always physical and hence ultimately quantum mechanical. However, what Landauer really claims is that nobody can have information without some physical carrier. People often confuse information with its carrier. From this confusion, such definitions as "information is a message" or "information is a collection of facts and data" or "information is knowledge derived from something" come.

However, if we identify information with its carrier, then we would inevitably come to the paradoxical conclusion suggested by Furner (2004) that information studies do not need the concept of information.

In reality, the situation is opposite. It is not the case when information laws are derived from physical laws, as Landauer (2002) suggested, but physical laws that are derived from information laws as the title of the book "Physics from Fisher information" (Frieden, 1998) states.

**Definition 6.** The substance *C* that is *physical carrier* of the portion of information *I* is called the *physical*, or *material*, *carrier* of *I*.

In a general case, carriers of information belong to three classes: *material*, *mental*, and *structural*. For example, let us consider a book. It is a physical carrier of information. However, it contains information only because some meaningful text is printed in it. Without this text it would not be a book. The text is the structural carrier of information in the book. Besides, the text is understood if it represents some knowledge and/or other structures from the cognitive infological system. This knowledge and other corresponding structures form the mental carrier of information in the book.

Two of the three types of information carriers are related to elements of the basic triplet (triad) of Jumarie (1986): the system *S* that is the *material medium* where information is physically defined; the *universe of discourse U* where information is semasiologically defined; and the *observer R* who considers *S* and *U* in his own subjective framework. The system *S* that is the material medium where information is physically defined corresponds to the physical carrier of information. The universe of discourse *U* where information is semasiologically defined corresponds to the structural carrier of information. The observer *R* who considers *S* and *U* in his own subjective framework corresponds to the mental carrier of information.

Distinctions between types of information carriers are of a great importance when something (such as a film or educational computer program) is produced for communication or/and entertainment. To achieve better information transmission, it is necessary to pay attention how all three types of information carriers are organized and produced.

Some people think that only physical matter is what gives emergent properties. They claim that with the same physical matter and with exactly the same physical structure, e.g., all microparticles and their states and coordinates, it is possible to get the same informational content. As a consequence, they believe, it is impossible to change informational content without changing the material representation of information.

This looks so evident. However, our practice shows that this is not the case. Let us consider a textbook on physics written in Japanese. To an individual who does not know either Chinese or physics, this book will give very little (if any) information. To an individual who knows physics but does not know Japanese, this book will give more information because this person will understand formulas. To an individual who knows Japanese but does not know physics, this book will give much more information. This person will be able to learn physics using this textbook However, to an individual who knows both Chinese and physics to a higher degree that this textbook represents, this book will also give very little (if any) information. Thus, the material representation of information in the book is not changing, while the content is different for different people. That is, what information is in the same material representation depends on means that the receiver of this information has for information extraction.

This conclusion is supported by statements of experts in statistical information theory where information content of a message depends on knowledge of the recipient/receiver.

Existence of a definite information carrier allows one to speak about this carrier as a *representation of information*. According to the Ontological Principle

O2, information is the same if it causes the same changes in a given infological system. Thus, the same information can be represented in different information carriers, e.g., by different texts, or even by information carriers of different nature, e.g., there cases when it is possible to convey the same information by an oral message, written or printed text, and picture.

However, there is a difference between information carrier and information representation. An information carrier only contains information, while an information representation contains and represents information. Thus, an information representation is always its carrier, while an information carrier is not always an information representation. An information carrier is a broader concept than a representation of information. For instance, a text written on a piece of paper is a representation of information and a carrier of this information as well. At the same time, the piece of paper with this text is only a carrier of the corresponding information. Note that one portion of information $I$ can represent another portion of information $J$. Thus, $I$ will be a carrier of $J$, but it will be a non-material carrier. A symbol is only partially a material carrier of information.

Here are some more examples of information carriers and information representations. A file that contains some text is both a representation and carrier of information, but the computer where this file is stored is a carrier but hardly a representation of information in the file. A human being is a carrier but not, as a rule, representation of information she or he has.

**Ontological Principle O4 (the Embodiment Principle).** *For any portion of information $I$, there is always a representation $C$ of this portion of information for a system $R$.*

The first three ontological principles ((O1)-(O3) or (O1)-(OM3)) imply that, in some sense, information connects the carrier $C$ with the system $R$ and thus, information is a component of the following fundamental triad (Burgin, 2004a):

$$(C, I, R) \qquad (5)$$

As a rule, there is some channel through which information comes from $C$ to $R$. For example, The carrier $C$ of $I$ is a piece of paper and $R$ is a person reading the text written on $C$. Then the corresponding channel is the space between the paper and the eyes of the person.

People empirically observed that for information to become available, the carrier must interact with a receptor that is capable to of detecting information the carrier contains. The empirical fact is represented by the following principle.

**Ontological Principle O5 (the Interaction Principle).** *A transaction/transition/transmission of information goes on only in some interaction of $C$ with $R$.*

This principle introduces the interaction triad (6) in the theory of information.

$$C \xrightarrow{Int} R \qquad (6)$$

Interaction between $C$ and $R$ may be direct or indirect, i.e. it is realized by means of some other objects.

The property of information explicated in the Ontological Principle O5 may look evident. However, it has important consequences, For example, if you know that some information has passed from system to another and want to find how it happened, you have to look for a channel of transaction. Although, if it is known only that the second system possesses the same information as the first one, it not necessary that it has been a transmission. It might be possible that the same information has been created by the second system. Existence of a channel makes transmission possible but does not necessitates it.

The next principle is a detailing of the Ontological Principle O5.

**Ontological Principle O5a (the Structured Interaction Principle).** *A system $R$ receives information $I$ only if some carrier $C$ of information $I$ transmits $I$ to the system $R$ or $R$ extracts this information from $C$.*

Information transmission/extraction can be direct or go through some channel *ch*. When somebody touches a hot rod and feels that the rod is hot, it is a direct information transmission from the rod to this person. At the same time, when somebody comes close to a hot rod and feels that the rod is hot, it is a indirect information transmission from the rod to this person because such a channel as the air is used.

Here, we have two ways of information transaction: transmission and extraction. Transmission of information is the passive transaction with respect to *R* when *R* receives information and active transaction with respect to *C* when *C* transmits information. Extraction of information is the active transaction with respect to *R* when *R* extracts information and passive transaction with respect to *C* when information is taken from *C*. When the carrier *C* is the system *R* itself, then we have the third type of information operations – information processing. It includes information transformation and production (Burgin, 1997b).

These two ways of information exchange reflect important regularities of education and entertainment media. At first, let us consider education where these features emerged much earlier than in entertainment.

There is an essential difference between Western and Eastern approaches to education. The main principle of the Western tradition is that a teacher comes to students to teach them. Contrary to this, the main principle of the Eastern tradition is that a student comes to teacher to learn from him. This means that the Western approach is based on information transmission, while the Eastern approach stems from information extraction.

This is an essential difference. When students come to school without preparing to work hard to get knowledge and only wait when the teacher will put everything in their head, the results usually are not good. Such students either do not receive knowledge or receive mush less than they can and what the teacher gives them. In fact, any teacher gives his students only information, while students themselves have to accept this information and to transform it to knowledge. This transformation demands substantial work. Gifted students do this work in their brains, often subconsciously. This creates an impression that they do nothing to achieve excellent results in learning. Not so gifted students need to work hard to achieve sufficient results in learning. It is a responsibility of a teacher to teach her students how to efficiently acquire information. Many problems with education, even in developed countries, are connected to the misconception that a teacher has to give knowledge to her students. This orients the teacher to give more subject material to students without taking sufficient care of helping the students to accept information in this material and to build knowledge using accepted information.

A similar situation, for example, exists in entertainment. To make it clear, let us consider theater, movies, television, and computer games. Theater evidently represents the Eastern position. Spectators come to the theater where plays are created. Film industry creates movies not in the presence of the audience. However, people have to come to movie theaters to see movies. TV, radio, and DVDs, as it is peculiar for the Western tradition, come to each home, making entertainment consumption easier. Computer games and other computer entertainment, in the majority of cases, are also coming to the audience. However, this audience consists not of the spectators but of participants. Modern technology provides for the film industry opportunities to come directly to the audience by means of the video. Thus, technology, mostly created in the West, supports and promotes active approach of producers at the cost of transformation of the audience (of spectators, students, etc.) into passive consumers. This is an essential trait of our society.

At the same time, the same Western technology has developed means for active participation in entertainment. Computer games represent only the first step in this direction.

The Ontological Principle O4a introduces the second communication triad (7) in the theory of information.

$$C \xrightarrow{\text{channel}} R \qquad (7)$$

Two more principles explicate dynamic properties of information processes.

**Ontological Principle O6 (Actuality Principle).** *A system **R** accepts a portion of information **I** only if the transaction/transition/transmission causes corresponding transformations.*

For example, if after reading this paper, your knowledge remains the same, you do not accept cognitive information from this text. That is why, the concern of the people from the entertainment industry how their production influences the intended audience is the greatest importance to the industry. General theory of information can explain many features of this impact. However, this theory does not solve all problems, and to have a complete picture, it is necessary to include sociologists, psychologists, economists, linguists, and semiologists in the study of entertainment.

**Ontological Principle O7 (the Multiplicity Principle).** *One and the same carrier **C** can contain different portions of information for one and the same system **R**.*

Really, let us consider some person *A* as the system *R* and a book written in Japanese as the carrier *C*. At first, *A* does not know Japanese and *C* contains almost no information for *A*. After some time, *A* learns Japanese, reads the book *C* and finds in it a lot of valuable information for himself. Note that knowing Japanese *A* is, in some sense, another person.

Another example is given by advertising. It changes comprehension of different thing including entertainment. Let us consider a situation when some person *A* comes to a movie theater to see a new film. It is, as a rule a great difference in comprehension depending whether this person never has heard about this film or *A* has read quite a deal about good actors, talented producer, interesting plot of the film and so on.

In other words, if you want to convey some information to an audience efficiently, you have to prepare this audience to acceptation of the transferred information. This is essentially important for contemporary entertainment industry based on mass communication.

There are many examples when unprepared community did not accept even the highest achievements of human intellect and creativity. Thus, it is known that when many outstanding works of art were created and many great discoveries in science were made, society did not understand what was done and rejected in some cases the highest achievements of geniuses. Only consequent generations understood the greatness of what had been done before. As examples, we can take the great Austrian composer Wolfgang Amadeus Mozart, who died in poverty, the great mathematicians Evariste Galois and Niels Hendrik Abel, who wrote outstanding works but were neglected and died at young age because of this.

One more example of misunderstanding gives the life of the great English physicist Paul Dirac. He was well-known and respected by physical community when he theoretically discovered a positive "electron", which was later called positron. However, other physicists did not understand Dirac's achievement and even mocked at him.

The great German mathematician Gauss made one of the most outstanding discoveries of the $19^{th}$ century, the discovery of the non-Euclidean geometry. Nevertheless, he did not want to publish his discovery because correctly considered the contemporary mathematical community unprepared to the comprehension of this discovery.

The last three principles reflect only the situations when transformation of an infological system takes place. However, it is important to know and predict properties of these transformations, for example, to evaluate the extent or measure of transformations. These aspects of the general theory of information are treated elsewhere.

It is true, to be sure, that an adequate theory, whether of information or anything else, must be in significant accord with our common ways of thinking and talking about what the theory is. Else there is the danger that the theory is not about what it purports to be about. Taking into account this aspect, we see that the general theory of information does not completely eliminate common understanding of the word information. This theory allows one to preserve common usage in a modified and refined form. Let us look how we change those expressions that are used as substitutes for the term *information* in the American

Heritage Dictionary (1996). The general theory of information suggests that it is more adequate to say and write that *information gives knowledge of a specific event or situation*. When people say and write that *information is a collection of facts or data* (The American Heritage Dictionary, 1996), the general theory of information suggests that it is more adequate to say and write that *a collection of facts or data contains information*.

According to the Dictionary, **Information** is: **1.** Knowledge derived from study, experience, or instruction. **2.** Knowledge of a specific event or situation; intelligence. **3.** A collection of facts or data: "statistical information." **4.** The act of informing or the condition of being informed; communication of knowledge: "Safety instructions are provided for the information of our passengers." **5.** (in Computer Science) A nonaccidental signal or character used as an input to a computer or communications system. **6.** A numerical measure of the uncertainty of an experimental outcome. **7.** (in Law) A formal accusation of a crime made by a public officer rather than by grand jury indictment.

According to the general theory of information, more adequate expressions are: **1.** Knowledge derived from information obtained from study, experience, or instruction. **2.** Information gives knowledge of a specific event or situation; or information provides intelligence. **3.** A collection of facts or data contains (statistical) information. **4.** The act of informing or the condition of being informed; communication. **5.** A nonaccidental signal or character used as an input to a computer or communications system contains information. **6.** A numerical measure of the uncertainty of an experimental outcome is a measure of information. **7.** A formal accusation of a crime made by a public officer contains information on who committed the crime.

It is possible to do similar transformations with the following definitions of information given in the Roget's New Thesaurus (1995): **1**. *That which is known about a specific subject or situation: data, fact* (*used in plural*), *intelligence, knowledge, lore.* **2**. *That which is known; the sum of what has been perceived, discovered, or inferred: knowledge, lore, wisdom.*

According to the general theory of information, more adequate expressions are: **1.** That which is known about a specific subject or situation, i.e., data, facts, intelligence, knowledge, and lore, contain information. **2**. That which is known contains information; the sum of what has been perceived, discovered, or inferred contains information, i.e., knowledge and lore contain information, while wisdom assumes possession of big quantity of information, i.e., a wise person has a lot of information.

This analysis of the common usage of the word *information* shows that the general theory of information does not essentially reverse the conventional meaning. The theory makes this meaning more precise by separating information from its representation. In our everyday speech, we do not differentiate between information and its representation. As a rule, it does not matter. However, this differentiation can be important in science and even in the every day communication, for example, when we need to find the intended meaning of a message, going beyond its literal understanding.

Principles of the general theory of information are introduced to reflect basic properties of information. But principles do not allow one to achieve the necessary exactness. That is why, mathematical structures are utilized, while principles are converted to postulates and axioms of the general theory of information are introduced. These postulates and axioms give an exact mathematical reflection of the main principles and provide for elaboration of a general axiomatic theory of information, which is based on the theory of named sets (fundamental triads). Fundamental triads are used for construction of the mathematical part of the general theory of information.